**Introduction**

A flock of birds circling over its roosting site is a magnificent aerial display. Theoretical work suggests that these highly synchronised and coordinated movements arise from simple interaction rules, without the need for centralised organisation (Vicsek et al. 1995; Couzin et al. 2002; Vicsek & Zafeiris 2012). Nonetheless, we are only just beginning to understand how rules implemented in models relate to those applied by animals. Progress in digital image processing and high temporal resolution tracking has allowed the inference of interaction rules in bird and fish species (e.g., Ballerini et al. 2008; Lukeman et al. 2010; Herbert-Read et al. 2011; Katz et al. 2011). Furthermore, in line with researchers' increasing interest in the role of inter-individual differences in shaping interactions (Conradt et al. 2009; Nakayama et al. 2012a), it has been found that homing pigeon flocks are hierarchically organised, where individuals contribute with different weights to the movement decisions of the flock (Nagy et al. 2010). Such hierarchical networks consist of transitive leader-follower relationships in which birds consistently copy the directional choices of individuals above them in the hierarchy, while being copied by those lower in rank. Little is known about what attributes of a flying pigeon can reliably predict leadership in flocks, although it has been suggested that leadership may be related to individual navigational efficiency (Nagy et al. 2010).

Empirical studies have identified a variety of traits (e.g., age, experience, social rank, and motivation; McComb et al. 2011; Reebs 2000; King et al. 2008; Nakayama et al. 2012b) that can modify an individual's behaviour towards conspecifics. With respect to the context of collective motion, recent work has demonstrated that navigationally less experienced birds are likely to follow more experienced conspecifics (Flack et al. 2012). More specifically, the larger the difference in homing experience between two partners, the higher the likelihood that the more experienced bird will emerge as leader. Additionally, in highly experienced birds the accuracy with which individuals recapitulate previously established idiosyncratic routes when flying solo has been suggested to predict relative influence when flying in pairs (Freeman et al. 2011), suggesting that some aspect of navigational certainty (or perhaps inflexibility) may promote leadership. These findings raise new questions about how variations in navigational knowledge possessed by individual members influence group dynamics in pigeon flocks. If a bird's position in the hierarchy correlates positively with its own navigational experience, we should be able to manipulate the network by providing selected individuals with the opportunity to acquire additional spatial knowledge. Here we evaluate whether it is indeed possible to alter individuals' ranks attained during flock homing flights by providing them with additional homing experience before re-testing them with their group mates.

**Methods**

*Subjects and experimental procedure*

We used 30 adult homing pigeons (*Columba livia*) bred at the Oxford University Field Station at Wytham (51°46'58.34''N, 1°19'02.40''W). They were kept in a social group of ca. 120 pigeons inside two lofts. Birds normally had free access to the outside, except on the days when the experiments were conducted. Food (a commercially available multigrain mixture), water, minerals and grit were provided ad libitum throughout the study. All experimental birds were between 4 and 8 years old, and had homing experience but had never visited the release site used in the current study. They were trained to carry miniature GPS logging devices (see below) attached to their back by a small Velcro strip glued to clipped feathers. All releases were performed from Radford (distance and direction to home: 15.7 km, 151°, respectively). The experiment had three phases. First, we trained three flocks of 10 birds (designated groups A, B and C), by releasing all 10 birds of a flock simultaneously at the release site (Phase I: group training). Each flock performed eight group training flights, with a maximum of two releases per day. We then calculated for each group a leadership hierarchy among flock members using the methods described in Nagy et al. (2010). In Phase II (solo training), we allowed three randomly chosen individuals from each flock to gain additional homing experience by performing 10 individual flights from the same site (one of these nine birds was lost during its 8th individual training flight, and therefore did not participate in the third phase for group C). Finally, in Phase III (group tests), we released each original flock six more times in order to evaluate any changes in the hierarchy's structure – in particular, whether the additional homing experience resulted in any changes in the ranks attained by the three individuals that had received additional solo training. Phase I was completed in 10 days, Phase II in 6 days, and Phase III in 3 days, with releases conducted on all consecutive days when weather conditions were favourable (dry and with winds<7 ms$^{-1}$).

*GPS device and data handling*

The GPS device was based on a commercially available product (Gmsu1LP, from Global Top), weighed 13 g, and was capable of logging time-stamped longitude, latitude and altitude data at 10 Hz. The geodetic coordinates provided by the GPS were converted into x, y and z coordinates using the Flat Earth model. These coordinates were smoothed by a Gaussian filter ($\sigma$=0.2 s), and we used a cubic B-spline method to fit curves onto the points obtained with the 0.1 s sampling rate. Only the x and y coordinates were used for analysis (the number of data points recorded per bird is shown in Table S1 of the Supplementary Material). In independent tests, using the devices in fixed relative positions to each other, the deviation between real and measured distance was 0.00 ± 0.34 m (mean ± S.D.). This degree of accuracy is sufficient for calculating directional correlation delay functions that characterise relations among the birds' motion (see Supplementary Figure S1 and Supplementary Methods for further details).

*Data analysis*

To evaluate the effect of training on homing performance, we calculated homing efficiency and homing time for each flight. Efficiency was measured by dividing the straight-line distance between the release site and the loft with the actual distance travelled by the bird to reach home. Homing time was the length of time that elapsed between release and the bird reaching a radius of 250 m from the loft. It should of course be noted that the two measures are not independent of each other, although the relationship between them can vary to some extent as a function of the bird's speed. In addition, to measure the trained birds' change in homing performance, we calculated the difference in efficiency and homing time between the average of the first two and the average of the last two solo training flights in Phase II.

To determine leader-follower relations inside the flock, we calculated the directional correlation delay for each pair of birds $i$ and $j$ ($i \neq j$). The directional correlation delay of a pair is $C_{ij}(\tau) = \langle \vec{v}_i(t) \cdot \vec{v}_j(t+\tau) \rangle_t$, where $\vec{v}_i(t)$ is the normalised velocity of bird $i$ at time $t$ and $\vec{v}_j(t+\tau)$ is the normalised velocity of bird $j$ at time $t + \tau$. Note that $C_{ij}(\tau) = C_{ji}(-\tau)$. We then determined the maximum value of the $C_{ij}(\tau)$ correlation function at $\tau_{ij}^*$, $C_{ij}(\tau_{ij}^*)$. We identified the corresponding $\tau_{ij}^*$ as the directional correlation delay time. $\tau_{ij}^*$ values focus on the relationship between specific pairings of individuals while ignoring hierarchy changes caused by other flock members. Note also that $\tau_{ij}^* = -\tau_{ji}^*$. Negative $\tau_{ij}^*$ values mean that the flight directional changes of bird $i$ fall behind that of bird $j$, and can thus be interpreted as a case of $j$ leading. In order to compare relationships among flock members before and after the solo training we focused on pairwise $\tau_{ij}^*$ values, averaged across pre- and post-training separately. For every specific pair $ij$, we averaged those $\tau_{ij}^*$ values that exhibited a $C_{ij}(\tau_{ij}^*)$ larger than 0.95. Because the relationships between specific pairings are non-independent data points, we used the number of individuals as our sample size for correlations between pre- and post- training $\tau_{ij}^*$ values. Only edges with values higher than 0.02 were retained.

For the calculation of the $C_{ij}(\tau)$ correlation function, we included only those pairs of data points from birds $i$ and $j$ where the two birds were a maximum of 100 m apart (i.e. $d_{ij}$<100 m). We chose this threshold based on the distributions of inter-individual distances (see Fig. S2).

A bird's closest neighbour was less than 10 m away in 71% of all recorded data points (see inset of Fig. S2). However, to be able to detect potential interactions between more distant flock members we used a threshold of 100 m, although only few data points fall into this bin category.

By averaging the $\tau_{ij}^*$ values of bird $i$ and the rest of the flock, we obtained a second measure, denoted $\tau_i$. Because of full transitivity of each hierarchy, this measure allowed us to resolve fully the hierarchical order among all group members. On two occasions, the $\tau_i$-value was 3.4 times higher than the standard deviation of all values (see Fig. S2); in these cases we removed the two birds from these particular flock flights and re-ran the analyses without them (see Table S2 in the Supplementary Material for the results including the outliers). We calculated for each bird the average of the $\tau_i$ values for the flights before (Phase I, 8 flights; $\bar{\tau}_i^{pre}$) and after (Phase III, 6 flights; $\bar{\tau}_i^{post}$) the individual training period. $\bar{\tau}_i$ values have similar properties to linear ranks (positive and negative values correspond to leading and following behaviour, respectively).

**Results**

Following the group releases of Phase I, we identified fully transitive hierarchies in each of our three flocks (Fig. 1, top row). Besides confirming the findings of Nagy et al. (2010), this initial result also provided the necessary premise for Phases II and III.

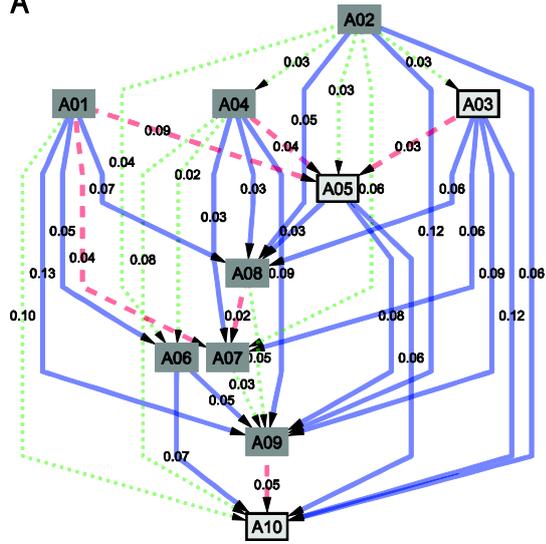
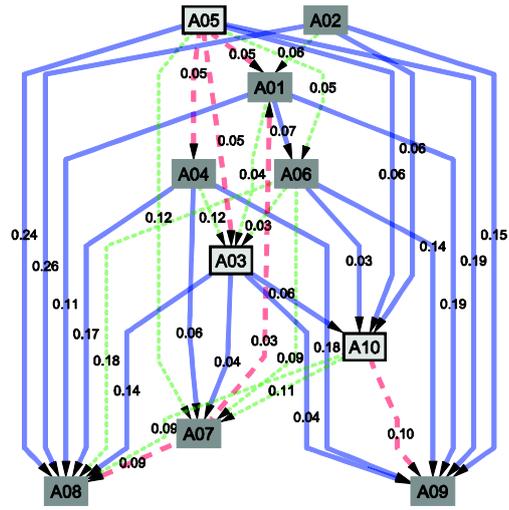
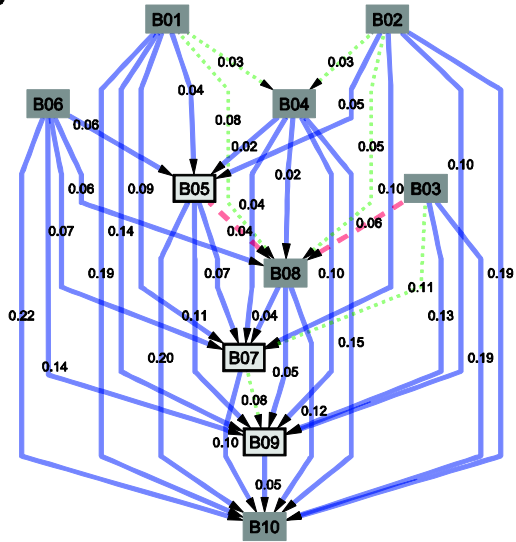
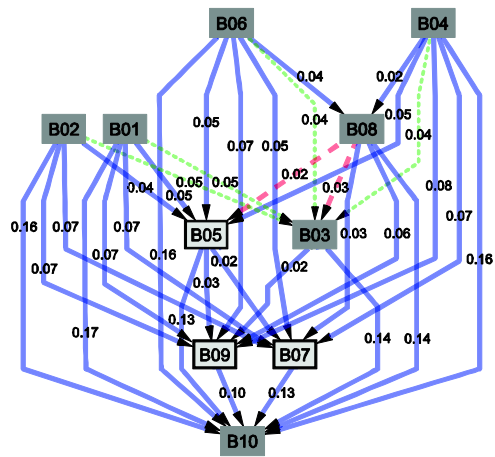
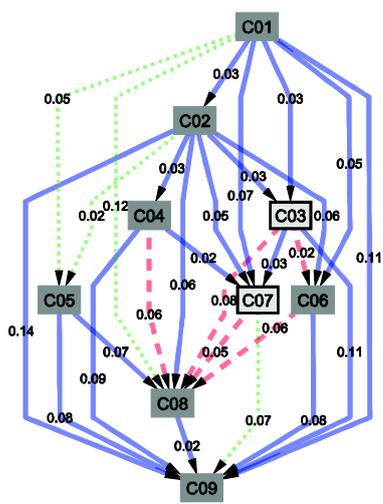
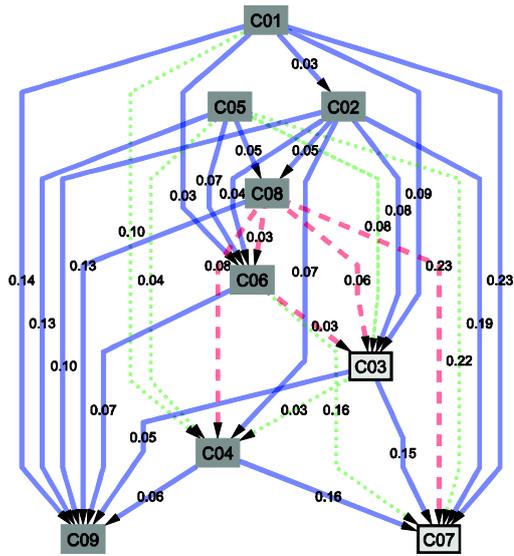

**Figure 1** Pre- and post-training hierarchical networks of three flocks, generated using $\tau^*_{ij}$ values. Rectangles correspond to individual birds; trained birds are shown with black borders. The three-digit alphanumeric codes indicate in which group the subject was tested (A, B or C) and its rank during the pre-training flights. Edges indicate leader-follower relations pointing from the leader to the follower (only edges where $\tau^*_{ij} \geq 0.02$ are shown). Edges that have the same directionality in pre- and post-training networks are indicated as thick blue lines; those that undergo a change in direction between pre- and post-training are shown as red lines; those that appear in only one of the networks are shown as dotted green lines. Numbers on edges correspond to $\tau^*_{ij}$. A, B, C Pre-training (left column) and post-training (right column) hierarchies of groups A, B and C, respectively.

First, we evaluated the effect of the training (flock and solo flights) on homing performance, by examining homing efficiency and homing time over the course of Phases I, II and III (Fig. 2A, B). Birds improved in both measures of homing performance during the flock releases of Phase I. Furthermore, during Phase II the solo trained birds increased their efficiency by an average of 0.13 (S.D.=0.06, difference between the average of the first two and the average of the last two solo training flights in Phase II, Fig. 2C) and decreased their homing time by -343.4 s (S.D.=209.0 s, Fig. 2D). Both these changes differed significantly from zero (one-sample t-tests, efficiency: $t_7$=5.77, P<0.001; time: $t_7$=4.65, P=0.002).

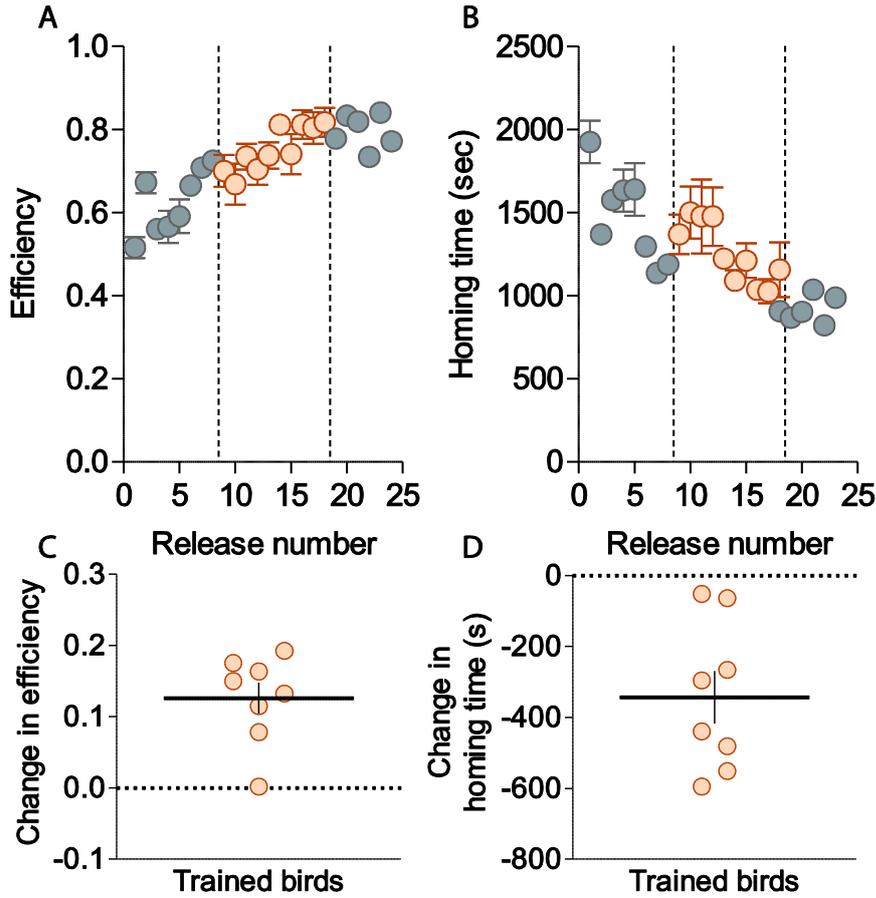

**Figure 2 (A, B)** Homing efficiency (mean ± S.E.M., **A**) and homing time (mean ± S.E.M., **B**) as a function of release number. Data from all groups were averaged according to release number. Grey circles indicate Phases I (N=30) and III (N=29), orange circles indicate Phase II (N=8). **(C, D)** Changes in homing efficiency **(C)** and homing time **(D)** during solo flights by trained individuals. Black line corresponds to mean (± S.E.M.).

We next used data from Phase III to measure the stability of the hierarchies by comparing the relative ranks of the untrained birds before and after solo training ($\bar{\tau}_i^{post}$ vs. $\bar{\tau}_i^{pre}$). We found a positive correlation between $\bar{\tau}_i^{pre}$ and $\bar{\tau}_i^{post}$ (Pearson's correlation, group ABC together: r=0.72, N=21, P<0.001, Fig. 3A, group A: r=0.80, N=7, P=0.030, group B: r=0.87, N=7, P=0.011, group C: r=0.69, N=7, P=0.090), which indicates the persistence of a robust hierarchical order among untrained flock members. However, the ranks of the trained birds exhibited variability: we found no correlation between $\bar{\tau}_i^{pre}$ and $\bar{\tau}_i^{post}$ (Pearson's correlation, r=-0.08, N=8, P=0.846, Fig. 3A), with some birds experiencing a rise and others a drop in $\bar{\tau}_i$. Also, the change in the birds' relative rank did not correlate with their changes in homing performance (Pearson's correlation, efficiency: r=0.247, N=8, P=0.556; time: r=0.072, N=8, P=0.866).

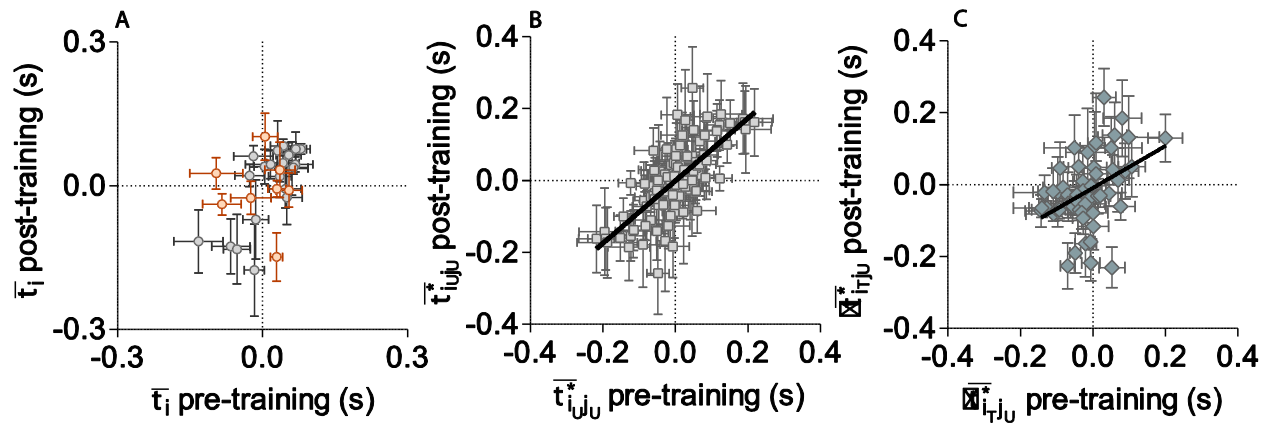

**Figure 3** Relationship between τ before and after individual training flights (mean ± S.E.M.) **(A)** $\bar{\tau}_i^{post}$ as a function of $\bar{\tau}_i^{pre}$ for solo-trained (orange circles) and untrained individuals (light grey circles). **(B-C)** averaged $\tau_{ij}^*$ after individual training as a function of averaged $\tau_{ij}^*$ before individual training for untrained-untrained pairings **(B)** and trained-untrained pairings **(C)**.

We further investigated the changes the hierarchies underwent using pairwise directional correlation time ($\tau_{ij}^*$). Again, we observed a positive correlation between pre- and post-training for the untrained birds (Pearson's correlation, group ABC together: r=0.72, N=21, P<0.001, Fig.3B, group A: r=0.70, N=7, P=0.083, group B: r=0.92, N=7, P=0.004, group C: r=0.62, N=7, P=0.138), further confirming the stability of their relationships over time and over repeated interactions. Moreover, pairwise $\tau_{i_T j_U}^*$ values enable us to compare the changes in pairs consisting of a trained ($i_T$) and an untrained ($j_U$) individual. The relationship between trained and untrained pigeons also remained stable, as evidenced by the positive correlation between their pre- and post-training $\tau^*_{i_T j_U}$ (Pearson's correlation, r=0.42, N=29, P=0.023, Fig. 3C). Despite the extra experience gathered by certain flock members, their positions in the hierarchy relative to untrained birds showed, on average, no improvement: the difference in directional correlation delay times in trained-untrained pairs before and after the individual training was on average 0.00 s (S.E.M.=0.01 s) and did not differ significantly from zero (one-sample t-test, $t_{55}$=0.005, P=0.996). Thus, although the overall hierarchical rank of the trained individuals changed slightly, the direction of these changes was not consistent. In addition, the changes were small enough that across the flock as a whole the position of the untrained birds in relation to trained flock members remained mostly unchanged. The extra training had an even smaller effect on the positions of the untrained birds relative to each other (Fig. 1). Separate examination of the three flocks showed that in group A two of the trained birds improved their relative ranks and one maintained its position (the average change in $\tau^*_{ij}$ before and after training, A: $\Delta\tau^*_{ij}$=0.06s (S.E.M.=0.02s), one-sample t-test, $t_{20}$=3.34, P=0.003, Fig. 1A). In group B, no clear change was found (B: $\Delta\tau^*_{ij}$=0.02 s (S.E.M.=0.01 s), one-sample t-test, $t_{20}$=1.77, P=0.09, Fig. 1B), whereas in group C the trained birds decreased their relative ranks C: ($\Delta\tau^*_{ij}$=-0.11 s (S.E.M.=0.02 s), one-sample t-test, $t_{13}$=5.30, P<0.001, see Fig. 1C). An additional statistical analysis, making use of the full dataset rather than per-bird averages as above, further confirmed the robustness of the measured hierarchies (see Supplementary Materials).

**Discussion**

Previous research has shown that group decision-making in pigeon flocks is hierarchically organised, with certain individuals consistently contributing with relatively more weight to movement decisions than others (Nagy et al. 2010). Here, we re-confirmed the existence of such hierarchical flight dynamics, demonstrating distinct leadership hierarchies in three separate flocks during repeated homing flights. Moreover, we showed that additional solo training given to specific group members did not affect the overall hierarchy of the flock: although trained birds increased their navigational efficiency during these solo flights (thus suggesting that they had gained additional navigational knowledge), this increase in efficiency was not accompanied reliably by improvement in their hierarchical position. Overall, pairwise leader-follower relations between flock members remained stable. This implies that leadership ranks within flocks do not directly relate to individual navigational experience, but that some other intrinsic property, or a combination of several properties, defines the organisation of the hierarchy.

Two possible mechanisms might allow the establishment and maintenance of robust flight hierarchies. The first requires individual recognition in the air. Flock members may have fixed leader-follower relationships that are based on individual identity and are consequently maintained across multiple flights. Such relationships may be similar to those based on dominance (King et al. 2008), familiarity (Flack et al. 2013) or individual affiliations (Jacobs et al. 2011). Alternatively, hierarchies might derive from individuals reacting in consistent ways to other group members' movements, without necessarily identifying them. Each individual may respond to flockmates in a way that is defined by its own specific features and the features it perceives in others. This would allow leadership to emerge passively as a consequence of simple interaction rules (Vicsek et al. 1995; Couzin et al. 2002). In other species these responses have been described to change with experience (Reebs 2000), motivation (Nakayama et al. 2012b), age (McComb et al. 2011), or sex (Ihl & Bowyer 2011).

The fact that we found no consistent effect of the extra training on birds' leadership ranks is a somewhat surprising result, given previous suggestions of the effect of navigational experience and skill on leadership (Nagy et al. 2010; Freeman et al. 2011). The trained birds' increase in experience might not have been large enough to induce changes in the organisation of the flock. Prior to the solo training, each subject had already performed eight flock homing flights and reached high, asymptotic levels of homing efficiency (Meade et al. 2005). Even though solo training did improve birds' solo homing efficiency, their advantage over the rest of the flock remained small or was only temporal. This interpretation is in agreement with past results showing that birds with more experience will more clearly emerge as leaders when the difference in experience between them and their flight partners is large (Flack et al. 2012). Future research should focus on the effect of experience while birds are still far from asymptotic levels of efficiency (e.g. with tests run after fewer homing flights for the most inexperienced birds). Furthermore, a control group in which every flock member receives extra solo training flights in Phase II would useful as a baseline measure of how flock homing efficiency changes in response to training given equally to all group members.

Flack et al. (2012) tested mixed-experience pairs of pigeons and found that navigational experience had an effect on leadership, with birds that had performed more training flights more likely to emerge as leaders. In the present study, using groups of ten birds, no such effect was detected, which may indicate that influencing flockmates' movements is easier in smaller groups. Also, theoretical work by Couzin et al. (2011) showed that the presence of uninformed individuals can inhibit decisions made by a knowledgeable minority and enable the numerical majority to control movements. Investigating the potential link between group size and group dynamics – both empirically and theoretically – is a promising avenue for future research.

Although flock dynamics can be observed without hierarchical organisation (Xu et al. 2012), such structure might be beneficial for establishing a "flight routine" that demands less attention from group members. The fact that hierarchies seem resistant to small changes once they are established indicates that rather than benefitting from particular features of the leader (such as navigational experience) their advantage might lie in the stability of the structure itself. Recent theoretical work has found that underlying social structures can improve the navigational accuracy of large, leaderless groups (Bode et al. 2012). Furthermore, it is suggested that hierarchical group dynamics could be based purely on social preferences (Bode et al. 2011). Social relationships can be found between relatives, familiar conspecifics or individuals of similar attributes such as size, personality or sex. Hence, the stability in our hierarchical networks may arise from preferential attachments that may have developed during training and that may not be susceptible to changes in individuals' navigational experience.

**Acknowledgements**

A.F. was supported by Microsoft Research, Cambridge. This work was partly supported by the EU ERC COLLMOT project (grant No. 227878) and EU ESF TÁMOP-4.2.1/B-09/1/KMR. M.N. was partly supported by a Royal Society Newton International Fellowship and Somerville College, Oxford. D.B. was supported by a Royal Society University Research Fellowship. The authors would like to thank Benjamin Pettit for technical assistance with the GPS tests and statistical advice. The authors are also grateful to three anonymous referees for helpful comments on an earlier version of the manuscript.

Supplementary Material for Flack et al., Robustness of flight leadership relations in pigeons

**Table S1** Mean ± S.D. number of data points per bird

| Group | A | B | C |
| --- | --- | --- | --- |
| Average no. of flock flight data points per bird | 167 725 (S.D.=7831) | 189 878 (S.D.=17422) | 169 082 (S.D.=8266) |

**Figure S1** Spatial and temporal error of the GPS trajectories and the directional correlation delay method for parallel (first column), globally fixed (second column) and perpendicular (third column) orientation tests. The pole (illustrated as coloured lines) was moved along a path (black line) in parallel (A), globally fixed (B) and perpendicular (C) orientation relative to the movement direction. **(D-F)** Relative position of each device in a pair relative to the direction of motion as a function of its actual position. D shows only one value of each pair ($i<j$). E and F show both values. **(G-H)** Probability density function (PDF) of the measured forward position of panel D-F. G shows the deviation between measured and actual position for each pair. **(J-K)** Forward ratio defined as the time ratio a device was detected to be at front relative to the motion direction. **(M-O)** Directional correlation function ($C_{ij}(\tau)$) between GPS 0 and all other devices. **(P-Q)** The directional correlation delay time ($\tau_{ij}$) of each pair.

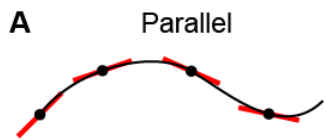
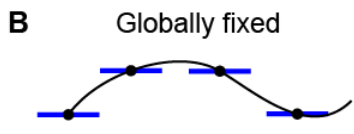
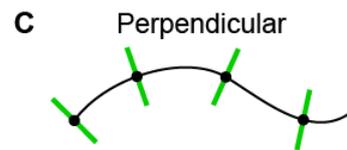
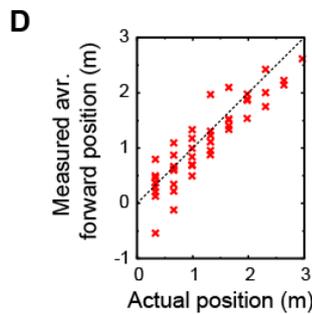
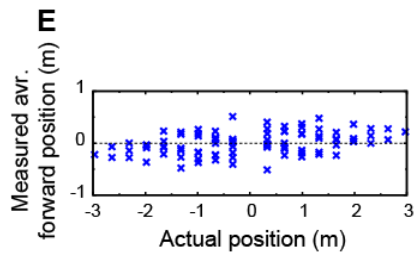
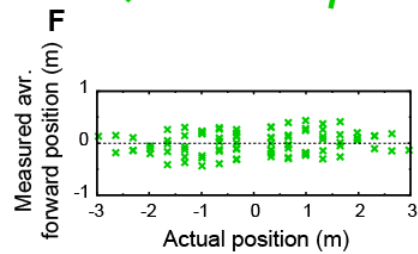
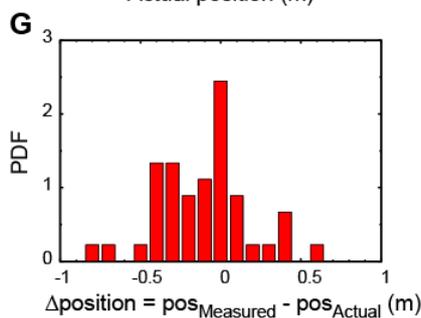
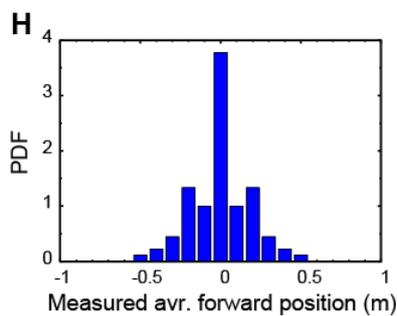
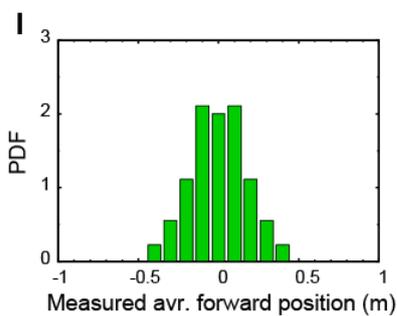
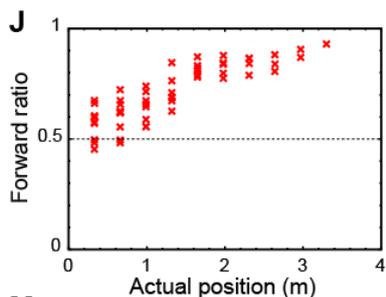
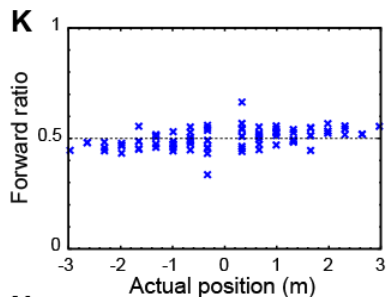
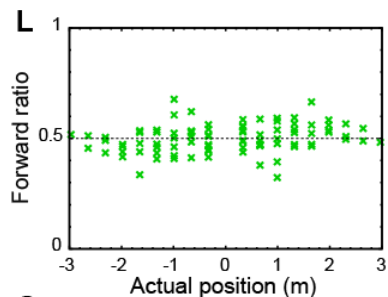
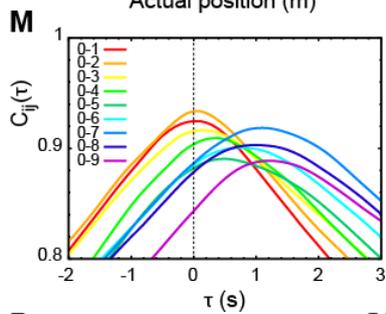
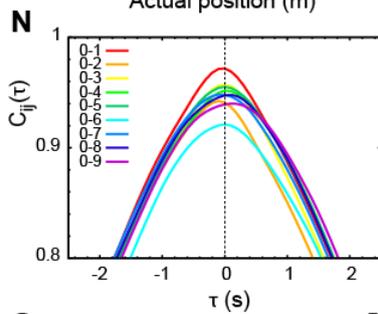
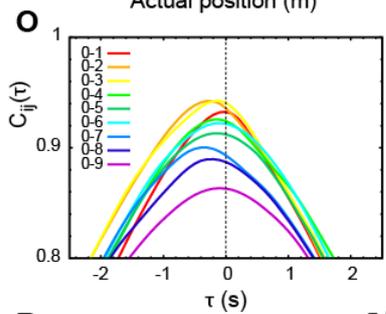
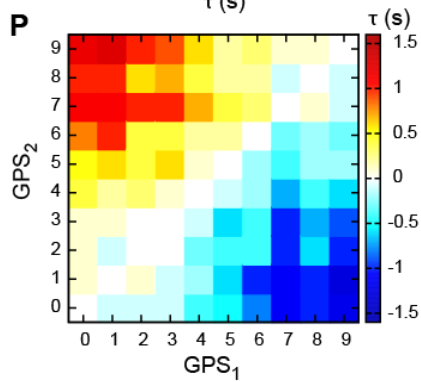
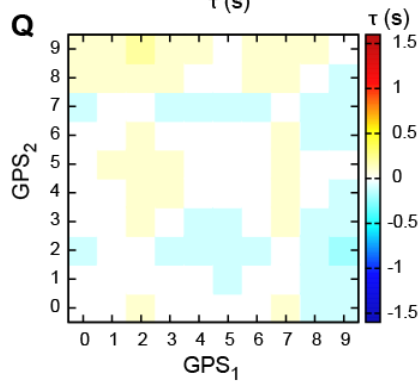
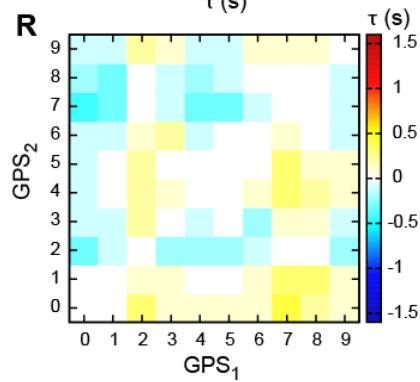

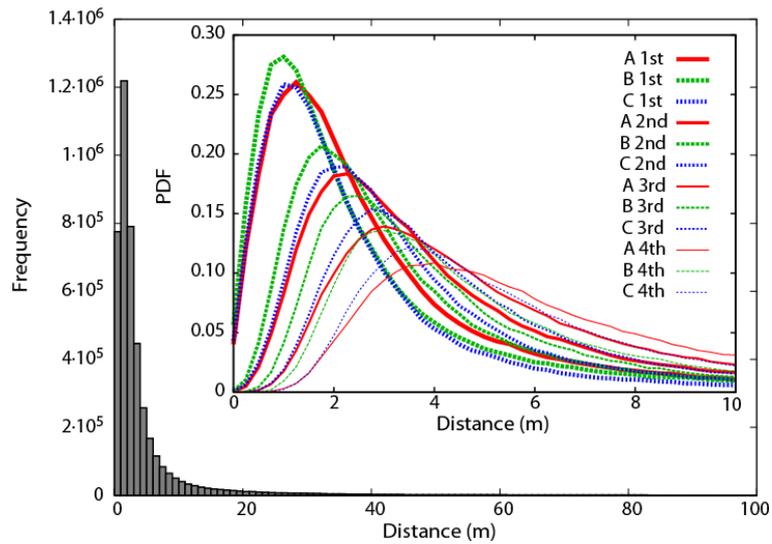

**Figure S2** Histogram illustrating the frequency distribution of distances (bin=1 m) to the first nearest neighbour of all three groups and flock flights (before and after solo training) pooled. Inset shows probability density functions of distances (bin=0.25 m) to the first, second, third and fourth nearest neighbours in groups A, B and C in red solid, green dashed and blue dotted lines, respectively (data shown only up to the fourth nearest neighbours for better visibility).

**Table S2** Results of the $\bar{\tau}_i^{pre}$ vs $\bar{\tau}_i^{post}$ correlation analysis with and without two outliers

| Correlation between $\bar{\tau}_i^{pre}$ and $\bar{\tau}_i^{post}$ | **Without outliers** | **With outliers** |
|---|---|---|
| untrained ABC (N=21) | **r=0.72, P<0.001** | **0.62, P=0.003** |
| untrained group A (N=7) | **r=0.80, P=0.031** | **r=0.80, P=0.031** |
| untrained group B (N=7) | **r=0.87, P=0.011** | r=0.61, P=0.149 |
| untrained group C (N=7) | r=0.69, P=0.090 | r=0.69, P=0.090 |
| trained ABC (N=8) | r=-0.08, P=0.846 | r=-0.176, P=0.677 |

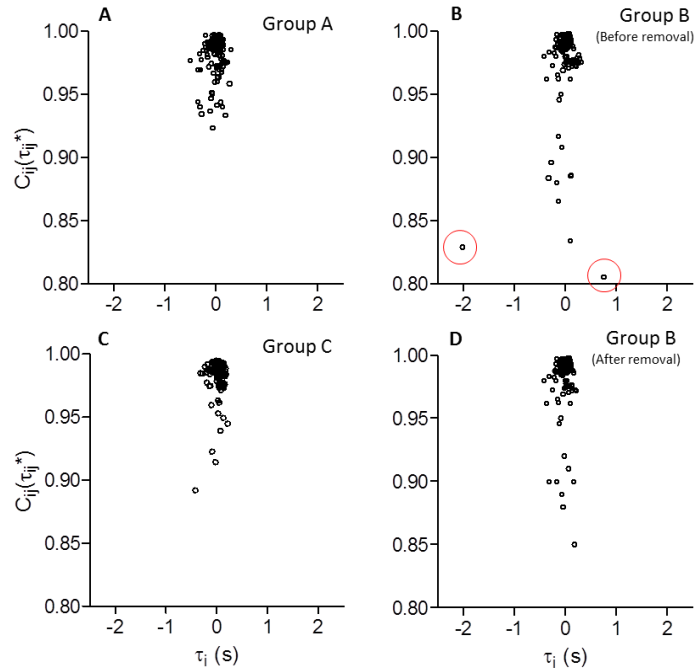

**Figure S3** Scatter plot of the relationship between an individual's $C_{ij}(\tau_{ij}^*)$ value and $\tau_i$ for groups A, B and C (panels (A), (B), and (C), respectively) for each flight. Red circles in B indicate $\tau_i$-outliers with low correlation values. Panel (D) shows the re-calculated $C_{ij}(\tau_{ij}^*)$, $\tau_i$ value pairs for group B after excluding those two outliers.

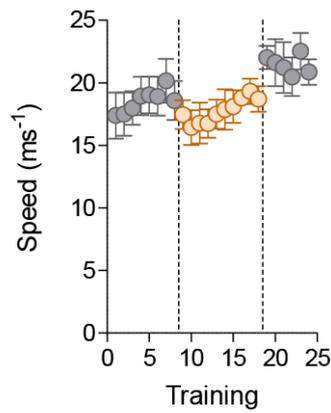

**Figure S4** Mean (± S.D.) speed as a function of training progression in Phases I, II and III. Data from all groups were averaged according to phase. Grey circles indicate Phases I (N=30) and III (N=29), orange circles indicate Phase II (N=8).

*Spatial and temporal error of the GPS devices and their impact on directional correlation delay analysis*

To test the spatial and temporal error originating from the GPS devices, we performed a variety of tests. 10 GPS devices (labelled 0 to 9) were attached to a rigid, 3 m long pole with an inter-device distance of 33 cm. We moved the pole along a free path in an open field using 3 different orientations: (1) with the pole's orientation parallel to the direction of motion (GPS 0 at the front and 9 at the back, Fig. S1A); (2) with the pole in a fixed orientation relative to the field (Fig. S1B); and (3) with the pole's orientation perpendicular to the direction of motion (Fig. S1C.) Each test lasted 10 minutes, and the pole moved between 1 and 3 ms$^{-1}$ (typical flight speed of a pigeons is 18-22 ms$^{-1}$).

An important aspect of analysing flock flights is the relative position of each device within a pair in relation to the movement direction of the whole flock. This is why we measured the average forward position of each device (Fig. S1D-E). In both, the perpendicular and the globally-fixed orientation case, we expect an average forward position of zero. We show the probability density function of this measure in Figure S1G-H. We also measured the time a device was detected to be in front relative to the direction of motion, and calculated the time ratio for the 10-min test (Fig. S1J-K). We also performed directional correlation delay analyses for all devices (Fig. S1 M-R). The absolute error of the GPS device arises from the relative error of the velocity which decreases as speed increases. Hence, our tests give an upper approximation of the noise due to the fact that each test lasted only 10 minutes and the pole was moved at low speeds.

*Additional test of hierarchy robustness*

We used a linear mixed-effects model to test the robustness of the hierarchies, using as our dataset the τ values calculated for each individual in every flight. All data were analysed using R (R Development Core Team, 2009) and the R packages lme4 (Bates & Maechler 2009) and languageR (Baayen 2009; cf. Baayen 2008). We included Subject as a random effect. As fixed effects, we added Training Phase (Phase I, pre-training or Phase III, post-training) and Treatment Group training group (trained or untrained individuals) to the model, as well as the interaction term between them.

We verified that the normality of error and homogeneity of variance assumptions of parametric analysis were statisfied by visual inspection of plots of residuals against fitted values. To assess the validity of the mixed effects analyses, we performed likelihood ratio tests comparing the models with fixed effects to the null models with only the random effect. The model that included fixed effects did not differ significantly from the null model (P=0.222). The given P-values were based on Markov-chain Monte Carlo sampling. The directional delay times did not differ from zero ($P_{MCMC}$=0.159), and we found no significant differences between pre- and post-training τ values ($P_{MCMC}$=0.656). The interaction between Training Phase and Treatment Group was not significant ($P_{MCMC}$=0.321). We also found no difference between pre- and post-training when examining untrained and trained birds in separate models (trained: $P_{MCMC}$=0.772; untrained: $P_{MCMC}$=0.219). Together, these results further confirm that the solo training had no effect on the groups' hierarchies.